
\input phyzzx

\def\pe{{\cal P}}
\def\to{\rightarrow}
\tolerance=500000
\overfullrule=0pt
\def\a{\alpha}
\def\l{\lambda}

\def\p{\partial}
\def\c{\prime}

\def\L{\Lambda}
\def\g{\gamma_{st}}
\def\k{\kappa}
\def\med{{1\over 2}}
\def\dkk{\pmatrix{2k\cr k\cr}}

\def\t{\theta}
\def\el{{\cal L}}
\def\ka{{\cal K}}
\def\D{\Delta}
\def\tm{t^{-1}}
\def\dkr{{R^k\over 4^k}\dkk}
\def\rma{\rho^\c_+(R)}
\def\rme{\rho^\c_-(R)}
\def\bloop{\tilde U(l)}
\def\floop{\tilde V_\pm(l)}
\def\bloop{\tilde U(p)}
\def\bloop{\tilde V_\pm(p)}
\def\sma{\sigma^+}
\def\sme{\sigma^-}
\def\smab{\sigma^{B+}}
\def\smeb{\sigma^{B-}}
\def\nma{\nu^+}
\def\nme{\nu^-}
\def\nmab{\nu^{B+}}
\def\nmeb{\nu^{B-}}
\def\lg{\langle}
\def\rg{\rangle}
\def\p{\partial}
\def\vp{V^\c}
\def\l{\lambda}
\def\u{{\widetilde U}}
\def\W{{\widetilde W}}

\pubnum={CERN-TH.6329/91}
\date={November, 1991}
\pubtype={}
\titlepage

\title{SUPERLOOP EQUATIONS AND\break
 TWO DIMENSIONAL SUPERGRAVITY}
\vskip 2.0cm
\author{L. Alvarez-Gaum\'e \break H. Itoyama \break
J.L. Ma\~nes \foot{Permanent
address: Departamento de F\'\i sica,
Facultad de Ciencias, Universidad
del Pa\'\i s Vasco, 48080 Bilbao, Spain}
\break and \break
A. Zadra}

\address{Theory Division, CERN\break
 CH-1211 Geneva 23, Switzerland}

\abstract{ We propose a discrete model whose continuum
limit reproduces the string susceptibility and the
scaling dimensions of $(2,4m)$-minimal superconformal
models coupled to $2D$-supergravity.  The basic
assumption in our presentation is a set of super-Virasoro
constraints imposed on the partition function.  We recover
the Neveu-Schwarz and Ramond sectors of the theory, and
we are also able to evaluate all planar loop correlation
functions in the continuum limit.  We find evidence to
identify the integrable hierarchy of non-linear equations
describing the double scaling limit as a supersymmetric
generalization of KP studied by Rabin.}

\endpage

\pagenumber=1

\def\ib{$$
\sum_{k\geq 1} kg_k{\p^{k-1}\over\p l^{k-1}}
w(l)=\int_0^l dl^\c w(l-l^\c)w(l^\c)
\eqn\eib
$$}

\def\iib{$$
Z=\int d^{N^2}\Phi \exp [-{N\over\L}tr V(\Phi)]
$$
$$
V(\Phi)=\sum_{k\geq 0}g_k\Phi^k
\qquad \L=e^{-\mu_B}
\eqn\eiib
$$}

\def\iiib{$$
w(l)={\L\over N}tr e^{l\Phi}=
\sum_{n=0}^{\infty}{l^n\over n!}{\L\over N}tr\Phi^n=
\sum_{n=0}^\infty {l^n\over n!}w^{(n)}
\eqn\eiiib
$$}

\def\ivb{$$
w^{(0)}=\L\;\;\;\;\; w^{(n)}=-\L^2 {\p F\over \p g_n}
\eqn\eivb
$$}

\def\vb{$$
Z(\mu_B)=\sum_h Z_h [N^2(\mu_B-
\mu_c)^{2-\gamma_{st}}]^{1-h}
\eqn\ev
$$}

\def\vib{$$
Na^{2-\g}=\k^{-1}
\eqn\evib
$$}

\def\viib{$$
L_n Z=0 \;\;\;\; n\geq -1
\eqn\eviib
$$}

\def\viiib{$$L_n={\L^2\over N^2}
\sum_{k=0}^n{\p^2\over \p g_{n-k}\p g_k}+
\sum_{k\geq 0}kg_k {\p\over\p g_{k+n}}
\eqn\eviiib
$$}

\def\ixb{$$
\L^2\sum_{k=0}^n{\p F_0\over\p g_k}
{\p F_0\over\p g_{n-k}}+\sum_{k\geq 1}kg_k
{\p F_0\over\p g_{k+n}}=0
\eqn\eixb
$$}

\def\xb{$$
w(p)=\int_0^\infty e^{-pl}w(l)dl=
\sum_{k=0}^\infty{w^{(k)}\over p^{k+1}}
\eqn\exb
$$}

\def\xib{$$
w(p)^2-V^\c(p)w(p)+Q(p)=0
$$
$$
Q(p)=\sum_{k\geq 1}kg_k
\sum_{r=1}^{k-1}p^{r-1} w^{(k-r-1})
\eqn\exib
$$}

\def\xiib{$$
\a_n=-{\L \sqrt 2\over N}
{\p\over\p g_n}\; , \; n\geq 0\;\;\; ; \;\;\;
\a_{-n}=-{N\over\L \sqrt 2}ng_n\; ,\; n>0
\eqn\exiib
$$}

\def\xiiib{$$
L_n=\med\sum :\a_{-k}\a_{k+n}:\;\;\; ,
\;\;\; T(p)=\sum_{n\epsilon Z} L_n p^{-n-2}
\eqn\exiiib
$$}

\def\xivb{$$
\eqalign{\p X(p)&=\sum_{n\epsilon Z}
\a_np^{-n-1}=\p X^++\p X^-\cr
\p X^+ &=\sum_{n\geq 0}\a_n p^{-n-1}\cr}
\eqn\exivb
$$}

\def\xvb{$$
\eqalign{
    \lim_{N\to\infty}{1\over N}Z^{-1}
\p X^+Z&\propto w(p)
    = \sum{w^{(n)}\over p^{n+1}}\cr
    \lim_{N\to\infty}{1\over N}Z^{-1}
\p X^-Z&\propto V^\c(p)=
    \sum kg_kp^{k-1}\cr}
\eqn\exvb
$$}

\def\xvib{$$
\lim_{N\to\infty} {1\over N^2} Z^{-1}T(p)Z=0
\eqn\exvib
$$}

\def\xviib{$$
Z={\rm const}\cdot\int\prod d\l_i
\prod_{i<j}(\l_i-\l_j)^2
\exp[-{N\over\L}\sum_i V(\l_i)]
\eqn\exviib
$$}

\def\xviiib{$$
Z={\rm const}\cdot\int\prod d\l_i
\Delta^2(\l)\exp[-{N\over\L}\sum_i V(\l_i)]
\eqn\exviiib
$$}

\def\xixb{$$
\sum_i\l_i^{n+1}{\p\Delta\over\p\l_i}=
\Delta\sum_{i\neq j}{\l_i^{n+1}\over
\l_i-\l_j}
\eqn\exixb
$$}

\def\xxb{$$
w(p)=\med\left( V^\c(p)-M(p^2)\sqrt{p^2-R}\right)
\eqn\exxb
$$}

\def\xxib{$$
w(p)={\L(R)\over p}+{w^{(2)}\over p^3}+
\ldots\;\;\; , \;\;\; w^{(0)}\equiv\L(R)
\eqn\exxib
$$}

\def\xxiib{$$
{\p w(p)\over\p\L}={1\over\sqrt{p^2-R}}
\eqn\exxiib
$$}

\def\xxiiib{$$
\L=e^{-\mu_B}=\L(R)=\sum_{k\geq 1}
kg_{2k}R^k\dkk {1\over 4k}=\oint_odxV^\c\left(
x+{R\over 4x}\right)
\eqn\exxiiib
$$}

\def\xxivb{$$
\eqalign{\L=1-(1-R)^m&=e^{-\mu_B}=1-a^2t\cr
              1-R    &=a^{2/m}u\cr}
\eqn\exxivb
$$ }

\def\xxvb{$$
{\p w^{(2k)}\over\p\L}={R^k\over 4^k}\dkk
\eqn\exxvb
$$}

\def\xxvib{$$
l=ka^{2/m}
\eqn\exxvib
$$}

\def\xxviib{$$
 w(l)\equiv\sqrt\pi\langle tr\Phi^{2k}
\rangle={1\over \k\sqrt l}\int_t^\infty
dt^\c e^{-lu(t^\c)}
\eqn\exxviib
$$}

\def\xxviiib{$$
\eqalign{
\langle w(l_1) w(l_2)\rangle &=
\pi\langle tr\Phi^{2k_1}
tr\Phi^{2k_2}\rangle\cr
&=-{\L\over N}\sqrt\pi{\p\over\p g_{2k_2}}\langle
 w(l_1)\rangle \;\; , \;\;
k_i=l_ia^{-2/m}\cr}
\eqn\exxviiib
$$}

\def\xxixb{$$
{\p u\over\p g_{2k}}=-a^{-2-1/m}
\sqrt{{l_2\over\pi}}e^{-l_2u}
\eqn\exxixb
$$}

\def\xxxb{$$
\langle  w(l_1) w(l_2)\rangle=
\sqrt{l_1l_2}{e^{-u(l_1+l_2)}\over l_1+l_2}
\eqn\exxxb
$$}

\def\xxxib{$$
\langle w(l_1)\ldots w(l_n) w(l_{n+1})
\rangle=-\k\sqrt{l_{n+1}}e^{-l_{n+1}u}
   {\p\over\p t}\langle w(l_1)
\ldots w(l_n)\rangle
\eqn\exxxib
$$}

\def\ic{$$
w(p,\Pi)\equiv v(p)+\Pi u(p)
\equiv\int_0^\infty dl
\int d\t e^{-pl-\Pi\t}w(l,\t)
\eqn\eic
$$}

\def\iic{$$
\eqalign{
\el [f]\equiv f(p,\Pi)&=
\int_0^\infty dz\int d\t e^{-pz-\Pi\t}
\left( f_0(z)+\t f_1(z)\right)\cr
 &=f_1(p)+\Pi f_0(p) \cr}\eqn \eiic
$$
$$\eqalignno{
f_i(p)&\equiv\int_0^\infty dz
e^{-pz}f_i(z)\;\; ,\;\;i=0,1 &(3.2a)\cr
\el [Df]=( \Pi +p &{\p\over\p\Pi})
\el [f]-f_0(0)\;\;\; ,\;\;\;
D={\p\over\p\t}+\t{\p\over\p z}&(3.2b)\cr
\el  \left[ ( \t-z{\p\over\p\t}) f\right]&=(
{\p\over\p\Pi}+\Pi
 {\p\over\p p}) \el [f] &(3.2c)\cr
\el [f\circ g] &=
-(-)^{\p f}\el [f]\el [g] &(3.2d)\cr
 (f\circ g)(z,\t)&\equiv\int
d\t^\c\int_0^z dz^\c f(z^\c,\t^\c)\, g(z-z^\c ,
\t-\t^\c)&\cr &=
f_1\circ g_0-f_0\circ g_1-
\t(f_1\circ g_1) &(3.2e)\cr}
$$}

\def\iiic{$$
\pe=\t-z{\p\over\p\t}
\eqn\eiiic
$$}

\def\ivc{$$
v(p)=\sum_{k\geq 0}{v^{(k)}\over p^{k+1}}
\;\;\; ,\;\;\;
u(p)=\sum_{k\geq 0}{u^{(k)}\over p^{k+1}}
\eqn\eivc
$$}

\def\vc{$$
T(p,\Pi)\propto :DX\p X:=
\psi\p_pX+\Pi:(\p_p X\p_p X+
\p_p\psi\psi ):
\eqn\evc
$$}

\def\vic{$$
\p X(p)=\sum_{n\epsilon Z}\a_n p^{-n-1}
\;\;\; ,\;\;\; \psi(p)=\sum_{r\epsilon Z+1/2}
 b_r p^{-r-1/2}
\eqn\evic
$$}

\def\viic{$$
\a_p=-{\L\over N}{\p\over\p g_p}
\;\; ,\;\; \a_{-p}=-{N\over\L}pg_p\; ,\;
p=0,1,2,\ldots \eqn\eviic
$$}

\def\viiic{$$
b_{p+1/2}=-{\L\over N}{\p
\over\p\xi_{p+1/2}}\;\; ,\;\; b_{-p-1/2}=
-{N\over \L}
    \xi_{p+1/2}\; ,\; p=0,1,2,\ldots
\eqn\eviiic
$$}

\def\ixc{$$
V(p,\Pi)=\sum_{k\geq 0}
(g_kp^k+\xi_{k+1/2}\Pi p^k)
\eqn\eixc
$$}

\def\xc{$$
u^{(0)}=\L\;\; ,\;\; u^{(n)}=-\L^2{\p F\over \p g_n}
\;\; ;\;\;\;\; v^{(n)}=-\L^2{\p F\over\p\xi_{n+1/2}}
\eqn\exc
$$}

\def\xic{$$
\lim_{N\to\infty} N^{-2}Z^{-1}T(p,\Pi)Z=0
\eqn\exic
$$}

\def\xiic{$$
KDw+DKw=wDw+Q
$$
$$
K=DV=\sum_{k\geq 1}(\xi_{k-1/2}+
\Pi kg_k)p^{k-1}\;\; ,\;\;\;\; Q=Q_1+\Pi Q_0
$$
$$
\eqalign{
Q_0&=2\sum_{k\geq1}\sum_{j=1}^{k-1}
\{kg_ku^{(k-j-1)}+(k-1-j/2)
\xi_{k-1/2}v^{(k-j-2)}
\}p^{j-1}\cr
Q_1&=\sum_{k\geq1}
\sum_{j=1}^{k-1}\{kg_kv^{(k-j-1)}+
\xi_{k-1/2}u^{(k-j-1)}\}p^{j-1}}
\eqn\exiic
$$}

\def\xiiic{$$
\pe\ka w(l,\t)+2\ka\pe w(l,\t)=
(w\circ\pe w)(l,\t)
$$
$$
\ka\equiv\sum_{k\geq 1}(kg_k\p_\t+
\xi_{k+1/2})\p_l^{k-1}
\eqn\exiiic
$$}

\def\xivc{$$
V  (\l,\t)=\sum_{k\geq 0}
\sum_i(g_k\l_i^k+\xi_{k+1/2}\t_i\l_i^k)
\eqn\exivc
$$}

\def\xvc{$$
Z(g,\xi)=\int\prod_i d\l_i
d\t_i\Delta (\l,\t) \exp [-{N\over\L}V(\l,\t)]
\eqn\exvc
$$}

\def\xvic{$$
G_r={\L^2\over N^2}\sum_{s=1/2}^r{\p^2
\over\p\xi_s\p g_{r-s}}+
\sum_{s=1/2}^\infty
   \xi_s{\p\over\p g_{r+s}}+
\sum_{k=1}^\infty kg_k{\p\over\p\xi_{k+r}}
\eqn\exvic
$$}

\def\xviic{$$
\eqalign{
L_n &={\L^2\over 2N^2}\sum_{k=0}^n{\p^2
\over\p g_k\p g_{n-k}}+\sum_{k=1}^\infty
   kg_k{\p\over\p g_{k+n}}\cr
    &+{\L^2\over 2N^2}\sum_{r=1/2}^{n-1/2}
({n\over 2}-r){\p\over\p\xi_r}
   {\p\over\p\xi_{n-r}}+
\sum_{r=1/2}^\infty({n\over 2}+r)
\xi_r{\p\over\p\xi_{r+n}}\cr}
\eqn\exviic
$$}

\def\xviiic{$$
G_{n-1/2}=\sum_{k=0}^\infty
\xi_{k+1/2}{\p\over\p g_{k+n}}+
\sum_{k=0}^\infty kg_k
  {\p\over\p\xi_{k+n-1/2}}+
{\L^2\over N^2}\sum_{k=0}^{n-1}
{\p\over\p\xi_{k+1/2}}
   {\p\over\p g_{n-1-k}}
\eqn\exviiic
$$}

\def\xixc{$$
g_{n-1/2}=-\t\l^n{\p\over\p\l}+
\l^n{\p\over\p\t}
\eqn\exixc
$$}

\def\xxc{$$
l_n=-\l^{n+1}{\p\over\p\l}-
\med (n+1)\l^n\t{\p\over\p\t}
\eqn\exxc
$$}

\def\xxic{$$
\sum_i(-\t_i\l_i^n{\p\over\p\l_i}+
\l_i^n{\p\over\p\t_i})
\eqn\exxic
$$}

\def\xxiic{$$
\sum_{k=0}^{n-1}\l_i^k\l_j^{n-1-k}=
{\l_i^n-\l_j^n\over\l_i-\l_j}\;\;\; ,\;\; i\neq j
\eqn\exxiic
$$}

\def\xxiiic{$$
\sum_i\l_i^n(-{\p\over\p\t_i}+
\t_i{\p\over\p\l_i})\Delta=
\Delta\sum_{i\neq j}\t_i
{\l_i^n-\l_j^n\over\l_i-\l_j}
\eqn\exxiiic
$$}

\def\xxivc{$$
\Delta=\prod_{i<j}(\l_i-\l_j-\t_i\t_j)
\eqn\exxivc
$$}

\def\xxvc{$$
Z=\int\prod_i d\l_i d\t_i
\prod_{i<j}(\l_i-\l_j-\t_i\t_j)
\exp [-{N\over\L}V  (\l,\t)]
\eqn\exxvc
$$}

\def\xxvic{$$
w(l,\t)\equiv{\L\over N}
\sum_ie^{l\l_i+\t\t_i}
\eqn\exxvic
$$}

\def\xxviic{$$
\eqalign{a)\;\;\;\;  &
(u(p)-V^\c (p))^2+(v(p)-\xi(p))^\c(v(p)-
\xi(p))=\Delta_0\cr
b)\;\;\;\;   &(v(p)-\xi(p))(u(p)-
V^\c(p)) =\Delta_1\cr}
\eqn\exxviic
$$}

\def\xxviiic{$$
\eqalign{
\xi(p) &\equiv
\sum_{k\geq 0}\xi_{k+1/2} p^k\cr
V^\c(p)&=\sum_{k\geq 1}kg_kp^{k-1}\cr}
\eqn\exxviiic
$$}

\def\xxixc{$$
\eqalign{
\Delta_0 &=V^\c(p)^2+
\xi^\c(p)\xi(p)-Q_0\cr
\Delta_1 &=\xi(p)V^\c(p)-Q_1\cr}
\eqn\exxixc
$$}

\def\id{$$
V  (\l,\t)=\sum_{k\geq 0}\sum_i(g_{2k}
\l_i^{2k}+\xi_{k+1/2}\t_i\l_i^k)
\eqn\eid
$$}

\def\iid{$$
w(p)={\L\over p}+O({1\over p^2})
\;\;\; ,\;\;\; \L=\L(R)
\eqn\eiid
$$}

\def\iiid{$$
u(p)={\L\over p}+O({1\over p^2})
\eqn\eiiid
$$}

\def\ivd{$$
v(p)-\xi(p)={\D_1\over u(p)-V^\c(p)}
\eqn\eivd
$$}

\def\vd{$$
u(p)-V^\c(p)=\sqrt{\D_0}-
{\D_1^\c\D_1\over 2\D_0^{3/2}}
\eqn\evd
$$}

\def\vid{$$
v(p)-\xi(p)={\D_1\over\sqrt{\D_0}}
\eqn\evid
$$}

\def\viid{$$
u(p)=V^\c(p)-M(p^2)\sqrt{p^2-R}+\ldots
\eqn\eviid
$$}

\def\viiid{$$
\eqalign{
v(p)&=\xi_+(p^2)+\xi_-(p)-N_-(p^2)
\sqrt{p^2-R}-{pN_+(p^2)\over\sqrt{p^2-R}}+
\ldots\cr
\xi_+(p^2)&=\sum_{k\geq 0}
\xi_{2k+1/2}p^{2k}\;\; ,\;\; \xi_-(p)=
\sum_{k\geq 0}\xi_{2k+3/2}p^{2k+1}\cr}
\eqn\eviiid
$$}

\def\ixd{$$
-{R\over M(R)} {n_0^+(R)(n_0^-(R)+
pn_1^+(R))\over (p^2-R)^{3/2}}
\eqn\eixd
$$}

\def\xd{$$
\eqalign{
M(R)&=M(p^2)\vert_{p^2=R}\cr
N_{\pm}(p^2)&=n_0^
\pm (R)+n_1^\pm (R)(p^2-R)+\ldots\cr}
\eqn\exd
$$}

\def\xid{$$
u(p)=V^\c(p)-M(p^2)\sqrt{p^2-R}-
{R\over M(R)}{n_0^+(n_0^-+pn_1^+)\over(p^2-R)^
   {3/2}}
\eqn\exid
$$}

\def\xiid{$$
v(p)=\xi_+(p^2)+\xi_-(p)-N_-(p^2)
\sqrt{p^2-R}-{pN_+(p^2)\over\sqrt{p^2-R}}
\eqn\exiid
$$}

\def\xiiid{$$
u_0(p)=V^\c(p)-M(p^2)\sqrt{p^2-R}
\eqn\exiiid
$$}

\def\xivd{$$
Z=\int\prod_{i=1}^{2N} d\l_i
d\t_i\prod_{i<j}(\l_i-\l_j-\t_i\t_j)
\exp [-{N\over\L}
\sum_k\sum_i g_{2k}\l_i^{2k}]
\eqn\exivd
$$}

\def\xvd{$$
u_0(p)={\L(R)\over p}+
{u^{(2)}\over p^3}+\ldots +
{u^{(2k)}\over p^{2k+1}}+\ldots
\eqn\exvd
$$}

\def\xvid{$$
\eqalign{
F(t)\equiv t^{-1/2}u_0(t^{-1/2})&=
g(t^{-1})-M(\tm)\tm \sqrt{1-Rt}\cr
g(p)&\equiv pV^\c (p)\cr}
\eqn\exvid
$$}

\def\xviid{$$
\eqalign{
M(\tm)&=\sum_{n\geq 0}
M_nt^{-n}\;\;\;  ,\cr
\oint_0 M(\tm)t^{l-1}dt&=
\oint_0 {g(\tm)\over\sqrt{1-tR}}t^ldt\cr}
\eqn\exviid
$$}


\def\xviiid{$$
{\p F(t,R)\over\p R}={1\over\sqrt{1-Rt}}
\Bigl(\med M(\tm)-{\p M(\tm)
\over\p R}(\tm-1)
\Bigr)
\eqn\exviiid
$$}

\def\xixd{$$
{\p F(t,R)\over\p R}=
{\p_R\L(R)\over\sqrt{1-Rt}}
\eqn\exixd
$$}

\def\xxd{$$
\p_R\L(R)=\med f(R)+R{\p f(R)\over\p R}
\eqn\exxd
$$}

\def\xxid{$$
\L(R)=\sum_{k\geq 1}kg_{2k}{R^k\over 4k}
\dkk=\oint_0dxV^\c(x+{R\over 4x})
\eqn\exxid
$$}

\def\xxiid{$$
M(p^2)\vert_{p^2=R}=2\p_R\L(R)
\eqn\exxiid
$$}

\def\xxiiid{$$
v_+(p^2)=\sum_{k\geq 0}{v^{(2k+1)}
\over p^{2k+2}}\;\;\; ,\;\;\;
v_-(p)=\sum_{k\geq 0}{v^{(2k)}\over p^{2k+1}}
\eqn\exxiiid
$$}

\def\xxivd {$$
v_+(p^2)=\xi_+-{pN_+\over\sqrt{p^2-R}}
\;\;\; ,\;\;\;
v_-(p)=\xi_--N_-\sqrt{p^2-R}
\eqn\exxivd
$$}

\def\xxvd{$$
v_+(\tm)=\xi_+(\tm)-{N_+
(\tm)\over\sqrt{1-Rt}}
\eqn\exxvd
$$}

\def\xxvid{$$
\oint_0\xi_+(\tm)\sqrt{1-Rt}\, t^{l-1}dt=
\oint_0 N_+(\tm)t^{l-1}dt
\eqn\exxvid
$$}

\def\xxviid{$$
{\p v_+(t,R)\over\p R}=-{1\over
(1-Rt)^{3/2}}\Bigl(\med N_+
(\tm)+{\p N_+(\tm)
\over\p R}(\tm-1) \Bigr)
\eqn\exxviid
$$}

\def\xxviiid{$$
\rho^\c_+(R)\equiv -\med N_{+0}+
R{\p N_{+0}\over\p R}=-\med\sum_{k\geq 0}
  \xi_{2k+1/2}{R^k\over 4^k}\dkk
\eqn\exxviiid
$$}

\def\xxixd{$$
N_+(p^2)=n_0^+ (R)+n_1^+ (R)(p^2-R)+\ldots
\eqn\exxixd
$$}

\def\xxxd{$$
n_0^+(R)=-2\rho_+^\c(R)
\;\;\; ,\;\;\; n_1^+(R)=-4\rho_+^{\c\c}(R)
\eqn\exxxd
$$}

\def\xxxid{$$
{\p u_0(p)\over\p\L}={1\over\sqrt{p^2-R}}
\;\;\; ,\;\;\;
   \L(R)=\sum_{k\geq 1}2kg_{2k}{R^k
\over 4^k}\dkk
\eqn\exxxid
$$}

\def\xxxiid{$$
{\p v_+(p)\over \p R}={p\rho^\c_+(R)
\over (p^2-R)^{3/2}}\;\;\; ,\;\;\;
   \rho^\c_+(R)=-\med\sum_{k\geq 0}
\xi_{2k+1/2}{R^k\over 4^k}\dkk
\eqn\exxxiid
$$}

\def\xxxiiid{$$
{\p v_-(p)\over\p R}={\rho^\c_-(R)
\over\sqrt{p^2-R}}\;\; ,\;\;
  \rho^\c_-(R)=\med\sum_{k\geq 0}(2k+1)
\xi_{2k+3/2}{R^k\over 4^k}\dkk
\eqn\exxxiiid
$$}

\def\xxxivd{$$
u_2(p)={R\over\p_R\L(R)}
{2\rho_+^\c(R)\bigl( \rho_-^\c(R)-
2p\rho_+^{\c\c}(R)\bigr)
 \over (p^2-R)^{3/2}}
\eqn\exxxivd
$$}

\def\xxxvd{$$
{\p u_0^{(2k)}(p)\over\p\L}=
{\p\over\p\L}\langle{\L\over N}
\sum_i\l_i^{2k}\rangle_0
     =\dkr
\eqn\exxxvd
$$}

\def\xxxvid{$$
{\p v^{(2k+1)}(p)\over\p\L}=
{\p\over\p\L}\langle{\L\over N}
\sum_i\t_i\l_i^{2k+1}\rangle =(2k+1)\dkr\rma\p_\L R
\eqn\exxxvid
$$}

\def\xxxviid{$$
{\p v^{(2k)}(p)\over\p\L}=
{\p\over\p\L}\langle{\L\over N}
\sum_i\t_i\l_i^{2k}\rangle=
\dkr \rme\p_\L R
\eqn\exxxviid
$$}

\def\xxxviiid{$$
u_2^{(2k)}=\langle{\L\over N}
\sum_i\l_i^{2k}\rangle_2=4k\dkr
\rma\rme\p_\L R\;\; ,
\;\; k\geq 1
\eqn\exxxviiid
$$}

\def\xxxixd{$$
u_2^{(2k+1)}=\langle{\L\over N}
\sum_i\l_i^{2k+1}\rangle=
-4(2k+1)\dkr\rma
 \rho_+^{\c\c}(R)R\p_\L R\
\eqn\exxxixd
$$}

\def\xld{$$
\eqalign{
\L(R)&=1-(1-R)^m+\sum_n t^B_n(1-R)^n\cr
     &=\L=e^{-\mu_B}\sim 1-a^2t\cr}
\eqn\exld
$$}

\def\xlid{$$
R=1-a^{2/m}u
\eqn\exlid
$$}

\def\xliid{$$
u^m-\sum t_n u^n=t
\eqn\exliid
$$}

\def\xliiid{$$
t_n^B=a^{2(1-n/m)}t_n
\eqn\exliiid
$$}

\def\xlivd{$$
\g=-1/m
\eqn\exlivd
$$}

\def\xlvd{$$
Na^{2+1/m}=\kappa^{-1}
\eqn\exlvd
$$}

\def\xlvid{$$
\smab_n=-{1\over 2(n+1)}
{N\over\L}\sum_{k=0}^{n+1}
{(-4)^k\over
  \dkk}\pmatrix{n+1\cr k\cr}
\sum_{i=1}^N\l _i^{2k}
\eqn\exlvid
$$}

\def\xlviid{$$
\lg\smab_n\rg=\lg \smab_n
\rg_0+\lg \smab_n\rg_2
\eqn\exlviid
$$}

\def\xlviiid{$$
\delta\L(R)=-{R\over n+1}
\p_R(1-R)^{n+1}=(1-R)^n-(1-R)^{n+1}
\eqn\exlviiid
$$}

\def\xlixd{$$
\sma_n=a^{2(1-n/m)}\smab_n
\eqn\exlixd
$$}

\def\ld{$$
\p_t\lg\sma_n\rg_0=
-{1\over 2\k^2}{u^{n+1}\over n+1}
\eqn\eld
$$}

\def\lid{$$
\lg\sma_n\rg\sim t^{1+1/m+n/m}
=t^{2-\g+(d_n-1)}
\eqn\elid
$$}

\def\liid{$$
d_n={n\over m}
\eqn\eliid
$$}

\def\liiid{$$
{\p \rho_\pm\over\p R}=
\mp\med\sum_n\tau_n^{B\pm}(1-R)^n
\eqn\eliiid
$$}

\def\livd{$$
\nmab_n=-{N\over\L}
\sum_{k=0}^{n}{(-4)^k\over
  \dkk}\pmatrix{n\cr k\cr}
\sum_{i=1}^N\t_i\l _i^{2k}
\eqn\elivd
$$}

\def\lvd{$$
\nmeb_n=-{N\over\L}
\sum_{k=0}^{n}{(-4)^k\over
  (2k+1)}{\pmatrix{n\cr k\cr}
\over\dkk}\sum_{i=1}^N\t_i\l _i^{2k+1}
\eqn\elvd
$$}

\def\lvid{$$
{\p\over\p\L}\Bigl({\L^2\over N^2}
\lg\nu_n^{B\pm}\rg\Bigr)
  =\rho^\c_\mp(R)\p_\L R(1-R)^n
\eqn\elvid
$$}

\def\lviid{$$
\smeb_n=-{N\over\L}\sum_{k=0}^{n}
{(-4)^k\over
  (2k+1)}{\pmatrix{n\cr k\cr}\over\dkk}
\sum_{i=1}^N\l_i^{2k+1}
\eqn\elviid
$$}

\def\lviiid{$$
\vp(p)=\sum_{k\ge 1}kg_kp^{k-1}
\equiv\vp_+(p)+\vp_-(p^2)
\eqn\elviiid
$$}

\def\lixd{$$
\vp_-(p^2)=\sum_{k\geq 0}(2k+1)g_{2k+1}p^{2k}
\eqn\elixd
$$}

\def\lxd{$$
u_0(p)=\vp_+(p)+\vp_-(p^2)-M(p^2)
\sqrt{p^2-R}-{pM_-(p^2)\over
  \sqrt{p^2-R}}+O(V_-^{\c 2})
\eqn\elxd
$$}

\def\lxid{$$
u(p)={\L(R)\over p}+
{\L_-(R)\over p^2}+ O({1\over p^3})
\eqn\elxid
$$}

\def\lxiid{$$
\p_R u_0={\p_R\L\over\sqrt{p^2-R}}+
{p\p_R\L_-\over(p^2-R)^{3/2}}
 \;\;\; ,\;\;\; M_-(p^2)
\vert_{p^2=R}=-2\p_R\L_-
\eqn\elxiid
$$}

\def\lxiiid{$$
\p_R \L_-=-\med
\sum_{k\geq 0}g_{2k+1}
{2k+1\over 4^k}\dkk R^k
\eqn\elxiiid
$$}

\def\lxivd{$$
\p_\L\Bigl({\L\over N}
\sum_i\l_i^{2k+1}\Bigr)=
{2k+1\over 4^k}
\dkk R^k{\p\L_-\over\p R}\p_\L R
\eqn\elxivd
$$}

\def\lxvd{$$
{\p\over\p\L}\Bigl
({\L^2\over N^2}\lg\smeb_n\rg\Bigr)
  =\p_R\L_-(R)\p_\L R(1-R)^n
\eqn\elxvd
$$}

\def\lxvid{$$
\p _R\L _-=-\med\sum_n t_n^{B-}(1-R)^n
\eqn\elxvid
$$}

\def\lxviid{$$
\lg\smeb_n\smeb_p\rg=
{\p\lg\smeb_n\rg\over\p t_p^{B-}}
$$
$$
{\p\over\p\L}\left( {\L^2\over N^2}
\lg\smeb_n\smeb_p\rg\right) =
\med(1-R)^{n+p}\p_\L R
\eqn\elxviid
$$}

\def\lxviiid{$$
\p_t \lg\smeb_n\smeb_p\rg=
-{1\over 2m\k^2}t^{1/m+n/m+p/m-1}
a^{2(n/m-1)}a^{2(p/m-1)}
\eqn\elxviiid
$$}

\def\lxixd{$$
\eqalign{
t_n^{B-}&=t_n a^{2(1-{n\over m})}\cr
\sme_n&=\smeb_n a^{2(1-{n\over m})}\cr}
\eqn\elxixd
$$}

\def\lxxd{$$
\lg\sme_n\sme_p\rg\sim t^{1/m+n/m+p/m}
\eqn\elxxd
$$}

\def\lxxid{$$
\eqalign{
{\L\over N}\lg\sum_i\l_i^{2k+1}
\rg&=-4R\rma\rho_+^{\c\c}(R)\p_\L R
{2k+1\over 4^k}\dkk R^k \cr
\lg\smeb_n\rg&=-4{N^2\over
\L^2}R \rma\rho_+^{\c\c}(R)\p_\L
R(1-R)^n\cr}
\eqn\elxxid
$$}

\def\lxxiid{$$
\lg\smeb_n\nmab_p\nmab_q\rg=
{N^2\over\L^2}(q-p)(1-R)^{p+q+n-1}
\p_\L R
\eqn\elxxiid
$$}

\def\lxxiiid{$$
\lg\smeb_n\nmab_p\nmab_q\rg=
{(q-p)\over m\k^2}
t^{(1/m-1)+n/m+p/m+q/m-1/m}
a^{2(n/m+p/m+q/m-1/m-3)}
\eqn\elxxiiid
$$}

\def\lxxivd{$$
\lg\prod_i O_i\rg\sim t^{2-\g+\sum (d_i-1)}
\eqn\elxxivd
$$}

\def\lxxvd{$$
\nma_n=\nmab_n a^{2(1-
{n\over m}+{1\over 2m})}
\eqn\elxxvd
$$}

\def\lxxvid{$$
d_p^++d_q^+={1\over m}(p+q-1)
\eqn\elxxvid
$$}

\def\lxxviid{$$
d_p^+={p\over m}-{1\over 2m}
\eqn\elxxviid
$$}

\def\lxxviiid{$$
{\p\over\p\L}\Bigl({\L^2\over N^2}
\lg\nmeb_q\rg\Bigr)
  =-\rma\p_\L R(1-R)^q
\eqn\elxxviiid
$$}

\def\lxxixd{$$
\p_t \lg\nmab_p\nmeb_q\rg=
-{1\over 2m\k^2}t^{(1/m-1)+p/m+q/m}
a^{2(p/m-1)+2(q/m-1)}
\eqn\elxxixd
$$}

\def\lxxxd{$$
\nme_n=\nmeb_n a^{2(1-{n\over m}
-{1\over 2m})}
\eqn\elxxxd
$$}

\def\lxxxid{$$
d_p^-={p\over m}+{1\over 2m}
\eqn\elxxxid
$$}

\def\lxxxiid{$$
\nu_n^\pm=\nu_n^{B\pm} a^{2(1-{n\over m}
\pm{1\over 2m})}\;\;\; ,
\;\;\; d_n^\pm={n\over m}\mp{1\over 2m}
\eqn\elxxxiid
$$}

\def\lxxxiiid{$$
\rho_\pm^\c=\mp\med
a^{2\pm{1\over m}}\tau^\pm(u)
\eqn\elxxxiiid
$$}

\def\lxxxivd{$$
\tau^\pm(u)=\sum_n \tau^\pm_n u^n \;\;\; ,\;\;\;
\tau_n^\pm=\tau_n^{B\pm}a^{-2(1-
{n\over m}\pm{1\over 2m})}
\eqn\elxxxivd
$$}

\def \ie{$$
\widetilde U(l) \equiv {\sqrt
{\pi \over l}} \sum _{i=1}^N\lambda
_i^{2k}\qquad l=ka^{2\over m}
\eqn \eie
$$}

\def \iie {$$
\widetilde U(l) = \u_0(l) + \u_2(l)
\eqn \eiife$$}

\def \iiie {$$
\lg\widetilde U_0(l)\rg=
{1\over \kappa l}\int _t^\infty
dt'\; e^{-lu(t')}
\eqn \eiiie
$$}

\def \ive {$$
u_2^{(2k)}=\left( \matrix {2k \cr
 k\cr }\right)  {k\over 4^{k-1}}\, \rho
_+'(R)\rho _-'(R)\, R^k\partial _\Lambda R
\eqn \eive
$$}

\def \ve {$$
\lg\widetilde U_2(l)\rg =
-{1\over \kappa }\, \tau ^+(u)\,
\tau ^-(u)\; e^{-lu}\partial _tu
\eqn \eve
$$}

\def \vie {$$
\lg\widetilde U(l)\rg={1\over \kappa l}
\int _t^\infty dt'\; e^{-lu(t')}-
{1\over \kappa}
e^{-lu}\partial _tu\; \tau ^+(u)\, \tau ^-(u)
\eqn \evie
$$}

\def \viie {$$
\widetilde V_+(l)\equiv 2\sqrt {\pi l}
\sum _{i=1}^N\theta _i\lambda _i^{2k}\qquad
\qquad \widetilde V_-(l)\equiv
\sqrt {\pi \over l} \sum _{i=1}^N
\theta _i\lambda
_i^{2k+1} \eqn \eviie
$$}

\def \viiie {$$
\langle \widetilde V_\pm (l)\rangle =
\pm {a^{1/m}\over \kappa}\int
_t^\infty dt'\; e^{-lu(t')}\, \tau ^\mp (u)\,
\partial _{t'}u
\eqn \eviiie
$$}

\def \ixe {$$
\W_\pm (l,\theta )\equiv \widetilde U(l) +
\theta _\pm ^B\widetilde V_\pm (l)
\eqn \eixe
$$}

\def \xe {$$
\theta _\pm =\theta _\pm ^B\,  a^{1/m}
\eqn \exe
$$}

\def \xie {$$\eqalign{
\partial _t \langle \sigma _n\rangle _0 &=
-{1\over 2\kappa ^2}\; {u^{n+1}\over
n+1}\cr
\langle \sigma _n\rangle _2&=
{u^n\over 2\kappa ^2}\; \tau ^+(u)\, \tau
^-(u)\, \partial _tu\cr}
\eqn\exie
$$}

\def \xiie {$$
\widetilde U(l)=2\kappa \sum _{n=0}^{\infty}
(-1)^{n+1}{l^n\over n!}\sigma _n +
({\rm singular\; as\;} l\to 0)
\eqn \exiie
$$}

\def \xiiie {$$
\partial _\Lambda \left(
{\Lambda ^2\over N^2}\, \langle \nu _n^\pm
\rangle \right) =-\rho _\mp '(R)\,
\partial _\Lambda R\, (1-R)^n
\eqn \exiiie
$$}

\def \xive {$$
\partial _t \langle
\nu _n^\pm \rangle =
\pm {u^n\over 2\kappa ^2}\; \tau ^\mp
(u)\, \partial _tu
\eqn \exive
$$}

\def \xve {$$
\widetilde V_\pm (l)=2\kappa
a^{1/m} \sum _{n=0}^\infty
{(-)^{n+1}\over n!}\, l^n\nu _n^\pm
+({\rm singular\; terms\; as\; } l\to 0)
\eqn \exve
$$}

\def \xvie {$$
\W_\pm (l,\theta )=\widetilde U(l)+
\theta _\pm ^B \widetilde V_\pm (l)=
2\kappa
\sum _{n=0}^\infty {(-)^{n+1}
\over n!}\, l^n (\sigma _n+
\theta _\pm \nu _n^\pm )
\eqn \exvie
$$}

\def \xviie {$$
{\rm dim\, }\nu _n^\pm =
{\rm dim \, }\sigma _n\mp {1\over 2m}
\eqn \exviie
$$}

\def \xviiie {$$
{\Lambda\over N}{\partial u
\over \partial g_{2k}}=
-2\k\sqrt {l\over \pi }\, e^{-lu}
\eqn \exviiie
$$}

\def \xixe {$$
{\Lambda \over N}\, {\partial \tau ^+
\over \partial \xi _{2k+{1\over 2}}}=\kappa
a^{1/m}\, {e^{-lu}\over \sqrt {\pi l}}
\eqn \exixe
$$}

\def \xxe {$$
{\Lambda \over N}\, {\partial \tau ^-
\over \partial \xi _{2k+{3\over 2}}}=2\kappa
a^{1/m}\, \sqrt {l\over \pi }\, e^{-lu}
\eqn \exxe
$$}

\def \xxie {$$
\langle \widetilde U(l_1)\cdots
\widetilde U(l_n)\rangle =
(-2\kappa \partial
_t)^{n-1}\langle \widetilde
U(l_1+\cdots +l_n)\rangle
\eqn \exxie
$$}

\def \xxiie {$$
\langle \widetilde V_\pm (l_1)
\widetilde U(l_2)\cdots \widetilde
U(l_n)\rangle =(-2\kappa \partial _t)^{n-1}
{\partial \over \partial \tau ^\pm
}\langle \widetilde U(l_1+\cdots +l_n)\rangle
\eqn \exxiie
$$}

\def \xxiiie {$$
\langle \widetilde V_+
(l_1)\widetilde V_-(l_2)
\cdots \widetilde
U(l_n)\rangle =(-2\kappa )^{n-1}
\partial _t^{n-3} {\partial \over \partial \tau
^+}{\partial \over \partial \tau ^-}
\langle \widetilde U(l_1+\cdots +l_n)\rangle
\eqn \exxiiie
$$}

\def \xxive {$$
{\cal D}_i^\pm \equiv
\partial _t + \theta _{\pm i}
{\partial \over \partial \tau ^\pm }
\eqn \exxive
$$}

\def \xxve {$$
\langle \W_+(l_1,\theta _1)
\cdots \W_+(l_{n_+}\theta _{n_+}) \W_-(l_{n_+
+1},\theta _{n_+ +1})
\cdots \W_-(l_n, \theta _n)\rangle =
\qquad \qquad \qquad \qquad $$
$$
\qquad \qquad \qquad
=(-2\kappa )^{n-1}\prod _{i=1}^{n_+}
{\cal D}_i^+ \prod _{j=n_+
+1}^n{\cal D}_j^- \quad \partial _t^{-1}
\; \langle \widetilde U(l_1+l_2+\cdots
+l_n)\rangle
\eqn\exxve
$$}

\def \xxvie {$$
\partial _t^{-1}\equiv - \int_t^\infty dt
\eqn\exxvie
$$}

\def \xxviie {$$
\langle \widetilde U(l)\rangle =
{1\over \kappa l}\left( -\partial _t^{-1} +\tau
^+\tau ^-\partial _t\right) e^{-lu}
\eqn\exxviie
$$}

\def \xxviiie {$$
\prod _i^{n_{_\pm }}{\cal D}_i^\pm =
\partial _t^{n_{_\pm }}+\theta _\pm \partial
_t^{n_{_\pm}\!\!\! -1}
{\partial \over \p\tau ^\pm }
\eqn\exxviiie
$$}

\def \xxixe {$$
\theta _\pm =
\sum _{i=1}^{n_\pm }\theta _{\pm i}
\eqn\exxixe
$$}

\def \xxxe {$$
-2\kappa ^2\, \partial _t^2 F=u-
\partial _t\left( \tau ^+\tau ^-\partial
_tu\right)
$$
$$
t=u^m-\sum _n t_nu^n
\eqn\exxxe
$$
$$
\tau ^\pm =\sum _n \tau _n^\pm u^n
$$}

\def \ifo {$$
\eqalign{ L&=D+u_1+u_2D^{-1}+u_3D^{-2}+...\cr
D&={\partial \over \partial \xi}+
\xi{\partial \over \partial x}\cr }
\eqn\eifo
$$}

\def \iif {$$
S=1+s_1D^{-1}+s_2D^{-2}+...
\eqn\eiif
$$}

\def \iiif {$$
S^{-1}LS=D
\eqn\eiiif
$$}

\def \ivf {$$
\eqalign{{\partial L\over
\partial t_{2n}}&=\left[ L_+^{2n},L\right]\cr
{\partial L\over \partial t_{2n-1}}&=
\left[ L_+^{2n-1},L\right] -2L^{2n}+\sum
_{k=1}^\infty t_{2k-1}
\left[ L_+^{2n+2k-2},L\right]\cr}
\eqn\eivf
$$}

\def \vf {$$
\eqalign {{\partial S\over \partial t_{2n}}&=
 - \left( S\partial _x^nS^{-1}\right)
_-S\cr
{\partial S\over \partial t_{2n-1}}&=
 - \left( S\partial _\xi
\partial _x^{n-1}S^{-1}\right) _-S\cr }
\eqn\evf
$$}

\def \vif {$$
w(z,\theta ,x,\xi, t)=z^{-1}S{\rm exp}
\left[ \sum _{n=1}^\infty \left(
 t_{2n}z^{-n} + t_{2n-1}\theta z^{-n+1}
\right) +xz^{-1} +\xi \theta \right]
\eqn\evif
$$}

\chapter{INTRODUCTION}
The discrete formulation of $N=1$
Superconformal Field Theories [1] coupled to
world-sheet supergravity lags far behind
its purely bosonic counterpart. Some of
the results in [2] were obtained previously
in terms of discrete models of
2D-gravity (see for example [3]). The supersymmetric
extension of the analysis
in [2] carried out in [4], [5] has no
analogues in terms of random
(super)-surfaces. In the double
scaling limit [6], the theory of the KP
(Kadomtsev-Petviashvili) hierarchy
was shown to play a central role [7].
Motivated by this connection, the authors in
[8] proposed an approach to the double
scaling limit of 2D-supergravity coupled
to superconformal matter using a
supersymmmetric generalization
of the KP hierarchy due
 to Manin and Radul [9]. An
important feature of the
 one-matrix model wich will
be central in our arguments is
the fact that its partition function satisfies
a set of Virasoro constraints [10],[11].

In this paper we propose a discrete model of
 2D-supergravity with superconformal
matter following the more ``phenomenological"
approach in [12].  Reasoning by
analogy with the Virasoro constraints we define
 a discrete analogue of the Hermitean
one-matrix model. Our basic postulate
 is to begin with a set of
super-Virasoro constraints in the Neveu-Schwarz (NS)
 sector satisfied by the
partition function. From them we derive the
 explicit form of our model and a set of
super-loop equations. In the planar limit
these equations can be solved exactly. In this
way we compute the spectrum of anomalous
 dimensions which coincide with the
(super)-gravitationally dressed dimensions of
 the $(2,4m)$ minimal superconformal
models in the NS and the Ramond (R) sectors.  We
compute arbitrary multiloop
 correlation functions on spherical
topologies, and our results agree with
those obtained in the continuum limit [13]
using the super-Liouville formulation
of the problem. Since
there are no higher point functions
computed in the continuum
we cannot compare our results further. In
 spite of these encouraging properties, a
derivation of our model in terms of
``triangulated super-surfaces", orthogonal
polynomials and generalized matrices is
 still lacking. Thus, the identification of our
model as a discrete version of 2D-supergravity
should be taken as preliminary.

We take as guiding principle in our work a set
of super-Virasoro constraints satisfied
by the partition function. One reason why we
 believe this to be a correct starting
point is the prominent role the Virasoro constraints
 play in the description of
the geometry of the moduli space of stable
 Riemann surfaces after Witten's work
[14] on 2D-gravity and matrix models, and
the proof of Witten's conjecture by
Kontsevich [15] (see also [16]). According
to Witten's theory, the
intersection theory of certain  line bundles
 on the moduli space ${\cal M}_{g,n}$ of
genus $g$ surfaces
with $n$ distinguished points is captured
 by the Virasoro constraints
satisfied by the partition function. Any discrete
 version of $N=1$ supergravity on the
world-sheet should necessarily have to
address similar issues for super-surfaces.
In this case, however, the mathematical
theory is not sufficiently well developed to
allow us to borrow results which could shed
light on the problem. We take a more
radical point of view in expecting many properties
 of super-moduli spaces to be
captured by the super-Virasoro constraints
$G_{n-{1\over 2}}Z=0$, $n=0,1,...$ . Since
$\{ G_n,G_m\} \sim L_{n+m}$, the Virasoro constraints
 are automatically
satisfied. The results of this paper can be
interpreted as giving support to the
validity of this basic assumption.

We also find evidence that the generalized KP
hierarchy appearing in our model is not
the super-KP hierarchy of Manin-Radul [9] but
 rather a super-hierarchy defined by
Rabin [17]. The basic difference between the two
 is related to the fact that the
latter does not admit a simple presentation in terms
 of a Lax pair. We will comment
on these issues at the end of the paper.

This paper is organized as follows. In section
 two we collect several results
concerning matrix models, loop equations and
Virasoro constraints and present them
in a way which will be easily generalized
later. In section three we study the
super-Virasoro constraints, we use them to
derive the exact form of our model and
then we derive the super-loop equation in
its planar approximation. Section four
analyzes the solution to the planar loop equations,
 the scaling limit and the spectrum
of scaling operators. We obtain the dressed
 gravitational dimensions expected in the
Neveu-Schwarz and Ramond sectors of the $(2,4m)$ minimal
 superconformal models
coupled to 2D-supergravity. As an appplication we
 compute in section five
correlators of arbitrary numbers of
loops (bosonic and fermionic) in planar
topologies. Section six contains our
 conclusions and out-look, and our remarks on the
connection between our model and the
 super-KP hierarchy described by Rabin [17].

\chapter{VIRASORO CONSTRAINTS AND LOOP EQUATIONS}
We review succintly in this section some properties of
the one-matrix models and
their loop equations. The starting point of Kazakov's
 analysis of multicritical points
[12] was the planar loop equation\foot{In the sequel we will
use the same symbol $w(l)$ for the loop operator and its
expectation value.  From the context it will be clear which
is the correct interpretation.}
\ib
$w(l)$ describes a loop of length $l$ bounding
a surface with the topology of a disk.
This equation follows from some simple heuristics,
but it can also be derived from a Hermitean
matrix model [18]. We take the partition function to be
\iib
where $\Phi$ is a Hermitean $N\times N$ matrix
and $\mu _B$ is the bare cosmological
constant. The loop operator is represented by
\iiib
Writing the partition function in terms
of the free energy $Z=e^{N^2F}$, $F=F_0 +
N^{-2}F_1+N^{-4}F_2+ ...$, the (expectation values)
 moments $w^{(n)}$ are given by
\ivb
Near the critical point $\mu _c$,
the genus expansion of \eiib\ behaves according to
\vb
where $h$ is the handle counting parameter
 and $\gamma _{st}$ is the string
susceptibility. Introducing the renormalized
cosmological constant $\mu _B-\mu_
c=a^2t$, where $a$ is a cut-off with units of
length, the double scaling limit [6] is
obtained by taking $N\to \infty$, $a\to 0$, and
 keeping fixed the combination
\vib
which is the string coupling constant. (More details
 and references to the literature
can be found in the review articles [19], [20]). For
later convenience we
derive the planar-loop equations \eib\
through the Virasoro constraints satisfied by
\eiib\ [21]. They are obtained by
making the change of variables $\Phi \to \Phi
+\epsilon \Phi ^{n+1}$. After some
simple manipulations we obtain
\viib
with
\viiib
To leading order in $1\over N$ (the planar limit)
we write $Z=e^{N^2F_0}$ and obtain
\ixb
Using \eivb\ it is easy to see that \eixb\ is exactly
identical to \eib\ . Hence  the
planar limit of the Virasoro constraint \eviib\
 is nothing but the planar loop equation.
The equations \eib\ \eixb\ are solved by introducing
the Laplace transform of the loop
operator
\xb
with the assumption that $w(l)$ behaves well
 at $l=0$ and $\infty $. With the
definition (2.10), (2.1) becomes an algebraic equation
\xib
Two remarks should be made at this point. First,
 if we make the identification
\xiib
the $L_n$'s can be rewritten as the components of
 the energy-momentum tensor of a
free massless scalar field
\xiiib
Defining
\xivb
we obtain
\xvb
and the Laplace transformed equation \exib\ becomes
\xvib
in the limit $p\to 0$. This is important
because with the identification \exvb\ we can
find the potential for the matrix
model. The planar loop equations are therefore
equivalent to \exvib\ . The second remark
 has to do with the form of $Z$ in terms of
eigenvalues. If the eigenvalues of
 $\Phi $ are $\lambda _1, ...,\lambda _N$, \eiib\ can
be written as [22]
\xviib
We could instead write
\xviiib
and use the Virasoro constraints \eviib\ \eviiib\
to determine the form of $\Delta $.
Indeed, if we act with \eviiib\ in \exviiib\
and perform some simple integrations by
parts, we obtain a set of differential
equations satisfied by $\Delta $
\xixb
whose solution up to an irrelevant constant
is $\Delta =\prod _{i<j}(\lambda
_i-\lambda _j)$. The solution to the planar
equations \exib\ takes the form [12]
\xxb
in the case when the potential is even $V(p)=V(-p)$. Since
\xxib
$M(p^2)$ is completely determined by
 requiring the right hand side of \exxb\ to have
only negative powers of $p$ in the large $p$
expansion, and $R$ is completely
determined in terms of
$\Lambda $. Kazakov showed [12] that
\xxiib
\xxiiib
At the $m$-th critical point
\xxivb
and $t$ and $u$ are scaling
 variables.  The variable $t$ is the
 renormalized cosmological
constant and $u$ is the ``heat capacity" of the theory.

To compute loop correlators notice from \exxiib\ that
\xxvb
Taking the limit as $k\to \infty $, and defining the
renormalized length according to
\xxvib
($l$ kept fixed as $a\to 0$, $k\to \infty $),
using the scaling limit \evib\ with $\gamma
_{st}=-{1\over m}$ and using the scaling
variables \exxivb\ we obtain
\xxviib
To compute multiloop correlation functions
 all we need to know is ${\partial u\over
\partial g_{2k}}$
\xxviiib
Perturbing the string equation \exxivb\ with
the $g_{2k_2}$ coupling and taking
$k_2\to \infty $ as in \exxvib\ we obtain
\xxixb
which together with \exxviib\ , \exxviiib\ yields
\xxxb
Repeating the same arguments we obtain
\xxxib
in agreement with the results in [23].

In the generalization to the supersymmetric case
in later sections we will follow
closely the arguments in this section.

\chapter{SUPER-VIRASORO CONSTRAINTS
AND SUPERLOOP EQUATIONS}
We now introduce the analogue of $w(l)$ which we
take to depend on two variables,
$w(l,\theta )$, a bosonic and a fermionic length
 $l$ and $\theta $ respectively. We
can imagine these two paramenters as characterizing
the boundary of a super-disk. As
in the previous case we can
introduce the super-Laplace transform
\ic
Some properties of the Laplace transform
 which are useful are the following:
\iic
where $\partial f$ is the grading of
$f$ ($\partial f=0$ if $f$ is even and $\partial
f=1$ if it is odd). In the last line $f_i\circ g_j$
is the standard convolution of
functions. We introduce also the symbol
\iiic
related to $D={\partial \over \partial
\Pi }+\Pi {\partial \over \partial p}$ via the
Laplace transform.

The first (of three) derivation of the planar
 loop equations is based on the analogy
with the $c=1$ energy-momentum tensor. Assuming again
that the loop $w(l,\theta )$
behaves well as $l$ is near $0$ or $\infty $, we can
expand the Laplace transform
$w(p,\Pi )$ in inverse powers of $p$
\ivc
Consider a  $\hat c=1$ free massless
 superfield $X(p,\Pi )= X(p) + \Pi \psi (p)$. Its
super-energy-momentum tensor is
\vc
Writing
\vic
we can identify $\alpha _n$, $b_r$ with
 bosonic and fermionic couplings
\viic
\viiic
The Laplace transformed loop $w(p,\Pi )$ is
identified with the positive frequency
part of $w(p,\Pi )\sim DX^+$ and the analogue of
the potential in the one-matrix model
is identified with $DX^-$, more
precisely $DV(p,\Pi )\sim DX^-$. This leads to
\ixc
Writing $Z=e^{N^2F}$ we can also identify
the moments $u^{(k)}$, $v^{(k)}$ with
derivatives of $F$:
\xc
With these identifications, we can take the planar
 limit as in the pure gravity case
\xic
with $Z\sim e^{N^2F_0}$. Some simple
algebra yields what should be considered as the
Laplace transform of the superloop equations:
\xiic
Using the Laplace transform formulae in (3.2) one
shows that \exiic\ is the Laplace
transform of the superloop equations
\xiiic
which is strongly reminiscent of \eib\
the starting point of Kazakov's analysis. It
should be quite interesting to derive \exiiic\
heuristically in terms of gluing
superdisks through their boundaries.

In the second and more fundamental
derivation of the superloop equations  (3.12),
(3.13) we begin by constructing an
``eigenvalue model" similar to \exviiib\ and use
super-Virasoro to determine the measure. The
form of the potential \eixc\ suggests
the introduction of $N$ pairs of
eigenvalues $(\lambda _i, \theta _i)$, one bosonic and
the other  fermionic. The potential for
this eigenvalue model is taken to be
\xivc
and the partition function is written as
\xvc
The explicit form of the super-Virasoro generators
in the $\hat c=1$ case with the
oscillators \eviic\ , \eviiic\ is given by
\xvic
\xviic
Since $L_n$ is obtained in
terms of anticommutators of $G_r$'s, it suffices to impose
on $Z$ only the fermionic constraints
$G_{n-{1\over 2}}Z=0$. It is convenient to write
$G_{n-{1\over 2}}$ as
\xviiic
If we recall the explicit
representation of the action of the algebra of supervector
fields (super-Virasoro without
central extension) on the space of functions of
$(\lambda ,\theta )$:
\xixc
\xxc
we find that the action of $G_{n-{1/2}}$ on $Z$ can
be traded off by the action of
\xxic
on the exponential term. Integrating
by parts and using identities like
\xxiic
we obtain a set of differential constraints on $\Delta $
\xxiiic
whose unique solution (up to an irrelevant multiplicative constant) is
\xxivc
Hence our model is explicitely given by
\xxvc
Introduce now the explicit representation of the superloop operator
\xxvic
and its expectation value with respect
to $Z$. Acting on $\langle w(l,\theta )\rangle $
with the operator $({\cal P}{\cal K}+2{\cal K}{\cal P})$
appearing in \exiiic\ , using the super-Virasoro
constraints and the factorization of
amplitudes in the large $N$ limit we obtain \exiiic\
after some computation.

Finally we could obtain \exiic\ , \exiiic\ by
using \exc\ , and the explicit formula for
$Z^{-1}G_{n-{1\over 2}}Z=0$, $n\geq 0$, and
$Z^{-1}L_nZ=0$, $n\geq -1$, with
$Z\sim e^{N^2F_0}$ and taking the large $N$
limit. This gives the third derivation of the
planar loop equations \exiic\ which we can write in
components according to
\xxviic
where
\xxviiic
and
\xxixc
$Q_0$ and $Q_1$ are given in \exiic\ although their
 explicit form is unnecessary
except for the fact that they are polynomials
in $p$. The primes denote differentiation
with respect to $p$.

After the three derivations presented of (3.27) the
 next step is to solve the
superloop equations, take the scaling limit and find
the critical points and critical
dimensions. This we do in the next section.

\chapter{SOLVING THE SUPERLOOP
 EQUATIONS. SPECTRUM OF THE MODEL}
In solving the superloop equations we
 will make the simplifying assumption that the
bosonic part of the potential is even
\id

Before attempting the solution of (3.27) there
are a number of useful remarks that
should be made concerning the solution of the
$2D$-gravity case (2.11), (2.20). As
written in \exib\ it seems that the
 polynomial $Q$ contains a number of ``initial
conditions" for some of the loop moments
$w^{(k)}$. If the potential is of order $k$,
apparently the first $k-1$ moments are completely
arbitrary, and all other moments
are computed in terms of them.

This is an incorrect impression if we
think in terms of the original matrix model.
Once the couplings $g_k$ and the cosmological
constant $\Lambda $ are given there
are no ambiguities. If we think in terms of the
formal perturbative evaluation, we
expand the potential in powers of all couplings
$g_k$ with $k\not =2$, i.e. we keep the
quadratic term $g_2\Phi^2$ in the exponent and
evaluate the free energy $F$ using
Wick contractions. The different correlators are
then formally analytic in all
couplings $g_k \; , \; k\neq 2$, and there are no
free constants. This can also be shown by
solving the Virasoro constraints perturbatively in
$g_k \; , \; k\neq 2$. This uniqueness
of the formal loop correlators when written in terms
of $(g_k,\Lambda )$ implies that
the solution to  \exib\ is unique provided: i) it is
parametrized by $\Lambda $ and it is
perturbative in the couplings $g_k \; , k\neq 2$;
ii) $w(p)$ only contains negative
powers of $p$ for $|p|$ large; and of course
iii) $w^2(p)-V'(p)w(p)$ is a polynomial,
which is another way of expressing \exib\ . Thus
although different ways of
parametrizing the cut in \exxb\ and the
function $M(p^2)$ might seem to yield
inequivalent answers, when we express
$R$ in terms of $\Lambda $ through the
condition
\iid
all the answers will be the same.

We can take the previous remarks and
 apply them in our situation. It is again true that
the perturbative evaluation of $Z$, or
the perturbative solution to the super-Virasoro
constraints yields unambiguous answers for
the superloop moments $u^{(k)}$,
$v^{(k)}$. Thus we stress the fact that if
 we find a superloop $w(p,\Pi )=v(p)+\Pi
u(p)$ satisfying: i) at large $|p|$ it only
contains inverse powers of $p$; ii) it is
perturbative in all couplings except the
quadratic even coupling; iii) the left hand
sides of (3.27) computed with the proposed
solution $w(p,\Pi )$ should be
polynomials ($\Delta _0$ and $\Delta _1$);
then when we use the string equation
\iiid
to express the auxiliary
parameters in terms of $\Lambda $, the solution is unique.
This uniqueness property of $w(p,\Pi )$
is very important and it will be used presently.
The solution may be parametrized differently
but the answers in terms of $\Lambda
$, $g_k$ $k\not =2$, $\xi _k$  will be the same.

To obtain a preliminay idea of the possible analytic
structure of $u(p)$, $v(p)$, we
solve (3.27) explicitely in terms of
$\Delta _0$, $\Delta _1$. The second equation in
(3.27) gives
\ivd
which turns the first equation in (3.27)
into a quartic equation in $u-v'$. However
since $\Delta _1$ is odd, $\Delta _1^2=0$,
 we obtain after some simple
manipulations
\vd
\vid
In principle $\Delta _0$, $\Delta _1$ are
 respectively even and odd polynomials in the
fermionic couplings $\xi _{k+{1\over 2}}$,
and a similar conclusion applies to $u(p)$
and $v(p)$. Among the two non-trivial solutions
to the quartic equation defining $u-V'$
we choose the one which gives a contribution to
 $u-V'$ of order zero in the fermionic
variables. Imitating the solution to the planar
bosonic case we choose $u(p)$ to have a
single cut
\viid
and for the fermionic operator $v(p)$
we take the starting ansatz
\viiid
{}From the form of \evd\ we easily guess that $u(p)$ must
 also have a contribution
proportional to $(p^2-R)^{-{3\over 2}}$. The polynomials
 $M$, $N_{\pm }$ are chosen to
cancel the positive powers in $g$,
$\xi _+$, $\xi _-$. Substituting \eviiid\ into the left
hand side of (3.27b) we see that we find a
 polynomial on the right hand side. As expected, with
\eviid\  we find that we need to
 modify the ansatz. It is easy to check
that it suffices to add to $u(p)$ the piece
\ixd
to guarantee that the left
 hand side of (3.27a) is a polynomial. In
\eixd\ $n_0^\pm (R)$,
$n_1^+ (R)$ and $M(R)$ are obtained by
expanding $N_\pm $, $M$ in powers of $p^2$
about $p^2=R$:
\xd
To summarize, our solution to the planar
superloop equations is given by
\xid
\xiid
As we will see below,
$N_\pm $ are linear in the fermionic couplings and $u(p)$ is at
most bilinear in fermions. This is
 rather surprising and one might be tempted  to add
higher order terms in the
$\xi $-couplings. Note however that (as soon as we give the
explicit form of $M$, $N_\pm $) $u(p)$ , $v(p)$
have only negative powers of $p$ as
$|p|\to \infty $, that the solution is perturbative
in the couplings $g_k$ $k\not =2$
and furthermore, by construction the left hand side of (3.27)
are polynomials. If the
reader has accepted the uniqueness arguments given
at the beginning of this section he
or she is unavoidably led to conclude that after
$R$ is traded by the cosmolgical
constant the same conclusion holds. Therefore any
other (more complicated)
dependence in the fermionic couplings is spurious
and it will be redefined away when
the correlators are expressed in terms of
$\Lambda $. This is purely a planar
phenomenon and it is expected not to hold to higher orders
in the large $N$ expansion in our
model \exxvc\ . We will offer a more geometrical
explanation of this fact in the next
section, after we evaluate correlation functions
of loop operators. Nevertheless this
feature of the planar superloop equations is rather
surprising.

The determination of $M(p^2)$ is exactly as
in the bosonic case, and this implies
therefore that the critical points will be
labelled by a positive integer $m$ with a
string susceptibility $\gamma _{st}=-{1\over m}$. Decompose
 $u(p)$ into
$u_0(p)+u_2(p)$, $u_0(p)$ (resp. $u_2(p)$) is of order
 zero (resp. two) in the fermionic
couplings
\xiiid
This form of $u_0(p)$ automatically implies the expansion
\xvd
We can arrive at the same result directly
from our model
 (when $V(p)$ is even) by showing
 that all odd expectation values
 $\langle \sum _i \lambda _i^{2k+1}\rangle$
vanish when all fermionic couplings are
 set to zero. In this case, the partition
function of our model takes the form
\xivd
The integration over
$\theta $'s yields $\Delta
(\lambda )Pf(\lambda _{ij}^{-1})$,
where $\Delta $ is the standard Vandermonde
determinant and $Pf(\lambda _{ij}^{-1})$
is the Pfaffian of the antisymmetric matrix
$M_{ij}=(\lambda _i -\lambda _j)^{-1}$
$i\not =j$, $M_{ii}=0$. Since $\Delta
(\lambda )Pf(\lambda _{ij}^{-1})$ is even under
$\lambda _i \to -\lambda _i$ it is clear
that any correlators of the form $\langle
\sum _i \lambda _i^{2k+1}\rangle $ vanishes.

Multiplying \exiiid\ by $p$ and introducing
the new variable $t=p^{-2}$ we have
\xvid
The left hand side $F(t)$ is
analytic near $t=0$: $F(t)=\Lambda (R) +u^{(2)}t+\ldots$.
  Similarly, for $t$ small
 $F(t)(1-Rt)^{-{1\over 2}}$ is also analytic near $t=0$,
hence $g(t^{-1})(1-Rt)^{-{1\over 2}}-M(t^{-1})t^{-1}$
should be analytic as well and we
can determine $M$
\xviid
We  need in particular
$$
M_0=\sum_{k=0}^\infty\dkk{2k+2
\over 4^k}R^k g_{2k+2}\equiv f(R)
$$
Furthermore
\xviiid
 by analyticity the term in parenthesis in \exviiid\ is
independent of $t$ and can
be computed using $f(R)$. From \exvd\ , \exvid\ we conclude
\xixd
\xxd
\xxid
Evaluating the term in parenthesis in
\exviiid\ (which is independent of $t$) at
$t^{-1}=R$, we obtain the useful identity
\xxiid

The determination of the fermionic
functions $N_\pm (p^2)$ goes
along the same lines.
Defining
\xxiiid
\exiid\ splits into two equations
\xxivd
The equation for $v_-(p)$ is identical
to the bosonic case. Introducing $t=p^{-2}$
\xxvd
since ${\sqrt {1-Rt}}\, v_+(t^{-1})$ is
 analytic near $t=0$, the analogue of \exvid\ becomes
\xxvid
and
\xxviid
As  $v_+(t,R)$ is divisible by $t$,
the term in parenthesis is independent of $t$, and
only the zeroth order in $N_+=\sum _{n=0}N_{+n}t^{-n}$ is relevant
\xxviiid
Expanding $N_+$ about $p^2=R$
\xxixd
and using the $t$-independence of the
term in parenthesis in \exxviid\ we obtain
\xxxd
We can carry out similar computations for
$v_-(p)$, and determine as well $u_2(p)$.
The final results can be summarized as follows:
\xxxid
\xxxiid
\xxxiiid
\xxxivd
Expanding in powers of $1\over p$,
we can write equivalently
\xxxvd
\xxxvid
\xxxviid
\xxxviiid
\xxxixd
The subscripts $0$ or $2$ in $\langle\ldots \rangle $
indicate orders $0$ or $2$
respectively  with respect to the fermionic couplings.
 Equations \exxxvd\ - \exxxixd\
are the basis for the computation of loop
correlators in section five.

To take the scaling limit of the previous expressions we
notice that the zeroth order
equations in fermionic couplings are those of the purely
bosonic theory.  Furthermore
$\Lambda (R)$ in \exxxid\ is independent of any fermionic
couplings. In analogy with
the analysis in [12], near the $m$-th critical point
$\Lambda (R)$ is given by
\xld
where $t$ is the renormalized cosmological
constant and $t_n^B$ is the bare coupling
of the $n$-th scaling operator.
Exactly at
the $m$-th critical point all $t^B_n=0$.  Introducing
the scaling variable
\xlid
The string equation and the renormalized couplings are
as in the bosonic case
\xliid
\xliiid
with  string susceptibility at the $m$-th critical point given by
\xlivd
and  in the double scaling limit we  again keep
\xlvd
fixed. From (4.35) and (4.40) we can
determine the form of the bare scaling
bosonic operators following Kazakov [12]:
\xlvid
There are two contributions to the planar expectation value:
\xlviid
Notice that the contribution to $\Lambda(R)$
of adding the operator (4.46) to the potential is
\xlviiid
and in the continuum limit only the first term
 survives.  This agrees	with
the identification of scaling operators in [24].  From (4.43)
the renormalized operator is
\xlixd
Since $\partial/\partial\Lambda =-a^{-2}\partial/\partial t$
we obtain
\ld
and
\lid
We can identify the gravitational dimension of $\sigma_n^+$ as
\liid
This result coincides with the scaling dimensions in the
NS-sector of a $(2,4m)$ $N=1$ superconformal
 minimal model coupled to $2D$-supergravity.

For the fermionic operators, in analogy with (4.40) we introduce the
fermionic scaling variables according to
\liiid
This definition determines the form of the fermionic operators
to be
\livd
\lvd
Hence
\lvid
To unambiguously determine the scaling behavior of the fermionic
variables $\tau^{B\pm}_n$, we find it convenient to work first
with the odd bosonic operators i.e. those determined by the	odd
expectation values $\sum\lambda_i^{2k+1}$.  From (4.39) we see that
we have three point functions involving one odd bosonic operator
and two fermionic operators $\nu^+$.  The explicit form of the odd
bosonic operators is
\lviid
To derive this expression we must go back to the original bosonic
loop equation (3.27a) and add a small odd perturbation to the bosonic
potential:
\lviiid
\lixd
The loop equation has a solution (to first order in $V_-'$):
\lxd
Following arguments similar to those explained previously, and
defining $\Lambda_-(R)$ by
\lxid
we obtain:
\lxiid
and
\lxiiid
Therefore
\lxivd
  We see that the operator (4.57)
behaves according to
\lxvd
\lxvid
We can now derive the scaling dimension of
$\sigma^{B-}_n$ by computing
two-point functions:
\lxviid
Hence
\lxviiid
and
\lxixd
Therefore the dimensions of the odd operators can be read off
from
\lxxd
 coinciding with the dimensions of the even operators
(4.52).  This duplication of operators is familiar from the
bosonic case.

The bosonic loop $u(p)$ in our solution contains a term bilinear
in $\rho_+$.  Equation (4.39) together with (4.57) implies:
\lxxid
{}From the scaling behavior of $\rho'_{\pm}$ (4.53) we obtain
\lxxiid
hence
\lxxiiid
Since for scaling operators we expect
\lxxivd
we conclude
\lxxvd
and
\lxxvid
We therefore identify
\lxxviid

Finally, from (4.36,37) and (4.54,55):
\lxxviiid
implying
\lxxixd
Thus
\lxxxd
and
\lxxxid
Summarizing
\lxxxiid
and these are the gravitational scaling dimensions of the operators in
the Ramond sector for the $(2,4m)$-minimal superconformal model
coupled to $2D$-supergravity.  Notice that in the spectrum (4.82)
all states are doubled except for the state with dimension
$d_0^-$.  This is a state in the boundary of the Kac table analogous to
the redundant operator $\sigma_{m-1}$ in the $m$-th critical bosonic
model. It is tempting to interpret $\nu_0^-$
 as related to the ground state of the theory.  We also
obtain the precise scaling behavior of $\rho'_{\pm}$:
\lxxxiiid
\lxxxivd
The reason why we had to go through
such a long argument to compute
the fermionic scaling dimensions is related to the extra
term $\pm 1/2 m$ in (4.82).  If
we had only considered (4.79)
the identification of the Ramond sector would have been
ambiguous.  To distinguish between
different possible assignments
we had to study carefully the coupling of $\sigma^-_n$ to
$\nu^+_p\nu^+_q$.  We have therefore shown that our model
in the planar limit has critical points labeled by
$m=1,2,3,\ldots$ and with scaling dimensions in agreement
with the NS- and R-sectors of the $(2,4m)$ superminimal models.
It is also easy to see that in the NS-sector the
one-, two- and three-point functions of scaling operators
agree with the results obtained in the continuum
super-Liouville theory [13].  We can compute arbitrary
correlators as well.  In the next section we compute the
correlation functions for an arbitrary number of
loop operators.

\chapter{PLANAR LOOP AND SUPERLOOP CORRELATION FUNCTIONS}
As an application of the result in the previous section
we compute correlation functions of loop operators.  We start
with bosonic loops.  We rescale for convenience by a factor
of $l^{-1/2}$ the definition of the loop operator in
(2.27).  This section is a translation of (2.25-31) to
the present situation.  Define
\ie
We will take $a\rightarrow 0$ and $k\rightarrow \infty$
keeping $l$ fixed.  There are two contributions in the planar
case to (5.1) coming from (4.35) and (4.38)
\iie
For $\u_0(l)$ the result is as given in (2.27) apart from a trivial
rescaling
\iiie
The other contribution comes from
\ive
Using (4.41,83) we obtain
\ve
In total
\vie
Similarly we can introduce two fermion loop operators
\viie
After some simple computations using (4.40,41,45,83) we arrive at
\viiie
It may seem surprising that $\langle
\tilde V_{\pm}(l)\rangle $ contains
a factor $a^{1/m}$.  However, when we think of superloops
with length $l$ and (bare) superlengths $\theta^B_{\pm}$:
\ixe
we define renormalized superlengths
\xe
and they behave dimensionally as $l^{1/2}$,
as we might expect.  It is possible to show that
\def\bloop{\tilde U(l)}
\def\floop{\tilde V_{\pm}(l)}
$\bloop$ and $\floop$ admit an
expansion in terms of microscopic
scaling operators.  The bosonic scaling operators (4.46) have
expectation values given by (4.50) and
$$
{\Lambda^2\over N^2}\langle \sigma^B_n\rangle_2=
-2[(1-R)^n-(1-R)^{n+1}]\rma\rme\partial_{\Lambda}R
$$
as one can show using (4.38).  In the continuum limit at the
$m$-th critical point:
\xie
This together with (5.6) yields
\xiie
The singular terms are analytic in $t$ and
they correspond to contributions
from microscopic loops.  Similarly in the fermionic case:
\xiiie
In the continuum limit
\xive
and comparing with (5.8) we obtain
\xve
Thus
\xvie
This result is dimensionally consistent
with the dimensions computed
for $\sigma_n,\nu^{\pm}_n$, $\widetilde
U\sim l^{-1/2}, \widetilde V_+\sim
l^{1/2},\widetilde V_-\sim l^{-1/2}$.
 Then we
use $\W_{\pm}=\bloop +\theta^B_{\pm}
\floop$ where $\theta^B_-$ is dimensionless but
$\theta^B_+\sim l^{-1}$.  Hence $[\theta_+]
\sim l^{-1/2}$ and $[\theta_-]\sim
l^{1/2}$.  Since $[l]=-1/m$ in units
of $t$, $[\theta_{\pm}]=\pm 1/2m$ and
\xviie
as expected.

Finally we compute multiloop correlation
functions.  We need to use equation (2.29),
and similar formulae for $\tau^{\pm}$.  It is
not difficult to show that when
$k\rightarrow \infty$ as in (2.26), \xviiie\
\xixe
\xxe
After some simple algebra one finds
\xxie
\xxiie
\xxiiie
{}From these expressions we can derive the
correlator of $n$ superloops
with $n^+$ $\W_+$ operators
and $n^-$ $\W_-$operators.  Define the
differential operators
\xxive
Then
\xxve
where
\xxvie
With this definition $\bloop$ in (5.6) becomes
\xxviie
Notice also that
\xxviiie
with
\xxixe
Therefore the superloop correlators
 (5.25) depend only on the total
length $L=l_1+l_2\ldots +l_n$ and the total superlengths
$\theta_{\pm}=\sum\theta_{\pm i}$.  A
posteriori this explains
why our solution to the loop equations
contains at most terms of
order two in the fermionic couplings.  This
is also reminiscent of
the bosonic case [25].  At the $m$-th critical
point the relevant
operators are $\sigma_0, \sigma_1,
\ldots \sigma_{m-2}$.  The operator
$\sigma_{m-1}$ is redundant and the
$L_{-1}$ Virasoro condition
in the scaling limit can be used to
compute correlators with one
insertion of $\sigma_{m-1}$.  In particular,
if in the planar limit
we evaluate $\langle \sigma_{m-1}
\prod w(l_i)\rangle$ we obtain simply
total length $L$ times $\langle
\prod w(l_i)\rangle$ and this
loop length counting operator
could be used to argue about the
dependence of planar multiloop
correlators on their total length.
We believe that similar arguments
should carry through in our case
but using instead the fermionic
partners of $\sigma_{m-1}$.

To conclude we should mention that the planar limit of the
theory is determined by the following set of equations
\xxxe
Rederiving the one- and higher-point functions from (5.30)
is left as an exercise to the reader.

\chapter{CONCLUSIONS AND OUTLOOK}
In our approach to the coupling of
minimal $N=1$ superconformal
models to $2D$-supergravity we
have taken as our guiding
principle a set of super-Virasoro
constraints satisfied by
the partition function (3.18).  This led to our explicit
representation of our model in (3.25).  From it we derived
the superloop equations and their
planar approximation (3.27), whose
solution led to a set of critical
 points labeled by an integer
$m=1,2,3,\ldots$ with string
susceptibility $\gamma _{st}=-1/m$
and with scaling operators
$\sigma_n, \nu_n^{\pm}$ with scaling dimensions
 identical to those expected in the NS- and R-sectors
of the $(2,4m)$-minimal superconformal
 model.  We were also
able to compute all multiloop
correlation functions.  There are
at least two outstanding problems with
regard to our model.  The first and
most important is the fact that
(3.25) is only an ``eigenvalue model"
with pairs of eigenvalues
$(\lambda_i,\theta_i)$ as basic
variables.  What kind of generalized
matrix model should lead to this
eigenvalue model is an important
open question, and we are quite
certain its solution will not
involve supermatrices [26].  The second
and easier problem
consists of proving the fact that
the superloop correlators
depend only on the total length and
superlengths by using the
properties of redundant fermionic operators.

A physically more interesting question
is to find out the properties
of supersymmetry breaking to all orders in the $1/N$-expansion
and non-perturbatively.  This requires finding the replacement
of the Painlev\'e-I equation found in [6] and the generalization
of Douglas's formulation [7] of the double scaling limit in terms
presumably of operators realizing a Heisenberg superalgebra.  The
resolution of this problem should also should shed some light on
how to generalize our model to the coupling of $(p,p')$-minimal
superconformal theories to $2D$-supergravity and also the explicit
form of the $\hat c=1$ theory.

The results presented in this paper lead us to believe that
the integrable system replacing the KP-hierarchy in our case is
not the super-KP-hierarchy constructed by Manin and Radul [9]
but rather the hierarchy (SKP) found by Rabin [17].  Both are
defined by starting with a pseudodifferential operator
\ifo
where we can identify $x$ with the cosmological constant
and $\xi$ with $\xi_{1/2}$.  Since $D^2=\partial/\partial x$,
$D^{-1}=\partial_x^{-1} D$.  Conjugating $L$ if necessary
$L\rightarrow hLh^{-1}$ in order to satisfy
$Du_1+2u_2=0$, there exists a
unique pseudifferential operator $S$
such that
\iif
satisfying
\iiif
The Manin-Radul hierarchy is obtained in terms of a Lax pair
constructed in terms of $L$.  If the variables
 $t_{2n}$ (resp. $t_{2n-1}$)
represent the even (resp. odd) flow parameters, and
$(L^n)_+$ denotes the differential part of $L^n$, the Manin-Radul
hierarchy is given by
\ivf
Note that in the odd flows we have an explicit dependence on
the odd parameters.  In Rabin's case the hierarchy is defined
according to the equations
\vf
where $(L^n)_-=L^n-(L^n)_+$.  This hierarchy is integrable but
it does not admit a representation exclusively in terms of $L$
as a Lax pair.  The wave function or Baker-Akhiezer function
takes the form
\vif
Identifying $\lambda$ with $z^{-1}$ and taking $\theta$ to be
the same as ours, we see that the expression in the exponential is
 the potential of our model.  We believe it should be possible
to show that our model is equivalent in the continuum limit to
this hierarchy or some minor  modification of it.  All these
questions are presently under investigation.

\vskip 1cm
{\bf ACKNOWLEDGEMENTS}.  We are grateful
to C. Gomez for fruitful
discussion on integrable supersymmetric
hierarchies and for bringing
reference [17] to our attention.  One of us (LAG) would like to
thank the Department of Theoretical
 Physics at the Universidad
Aut\'onoma de Madrid for their
hospitality while part of this
work was done. The work of JLM was
supported in part by CICYT and by a
UPV research grant (UPV 172.310-0151/89).
AZ is supported in part by the
World Laboratory.

\endpage

\noindent
REFERENCES

\vskip .4cm
\noindent
{\bf [1]} D. Friedan, Z. Qiu and
 S. Shenker, {\it Phys. Rev. Lett.} {\bf 51} (1984) 1575.

\noindent
{\bf [2]} V.G. Knizhnik, A.M. Polyakov
 and A.B. Zamolodchikov, {\it Mod. Phys. Lett.} {\bf
A3}
 (1988)  819.

\noindent
{\bf [3]} F. David, {\it Nucl. Phys.} {\bf B257} (1985) 45.

V. Kazakov, {\it Phys. Lett.} {\bf 150B} (1985) 282.

V. Kazakov, {\it Phys. Lett.} {\bf 119B} (1986) 140.

I.K. Kostov and M.L. Mehta,
{\it Phys. Lett.} {\bf 189B} (1987)118.

\noindent
{\bf [4]} A.M. Polyakov and
A.B. Zamolodchikov, {\it Mod Phys. Lett.}
{\bf A3} (1988)
1213.

\noindent
{\bf [5]} J. Distler, Z. Hlousek and H. Kawai,
{\it Int. J. Mod. Phys.} {\bf A5} (1990) 391.

\noindent
{\bf [6]} E. Br\'ezin and V.A. Kazakov,
{\it Phys. Lett.} {\bf 236B} (1990) 144.

M.R. Douglas and S.H. Shenker,
{\it Nucl. Phys.} {\bf B335} (1990) 635.

D.J. Gross and A.A. Migdal,
{\it Phys. Rev. Lett.} {\bf 64} (1990) 127.

\noindent
{\bf [7]} M.R. Douglas,
{\it Phys. Lett.} {\bf 238B} (1990) 176.

\noindent
{\bf [8]} P. Di Francesco,
J. Distler and D. Kutasov, {\it Mod. Phys. Lett.}
{\bf A5} (1990)
2135.

\noindent
{\bf [9]} Yu.I. Manin and A.O. Radul,
{\it Commun. Math. Phys.} {\bf 98} (1985) 65.

\noindent
{\bf [10]} R. Dijkgraaf, E. Verlinde and
H. Verlinde, {\it Nucl. Phys.} {\bf B328} (1991)
435.

\noindent
{\bf [11]} M. Fukuma, H. Kawai and
R. Nakayama, KEK preprint KEK-TH-251 (May
1990).

\noindent
{\bf [12]} V.A. Kazakov, {\it Mod.
Phys. Lett.} {\bf A4} (1989) 2125

\noindent
{\bf [13]} E. Abdalla, M.C.B. Abdalla,
D. Dalmazi and K. Harada, IFT preprint
 IFT-91-0351.

K. Aoki and E. D'Hoker, UCLA preprint UCLA-91-TEP-33.

L. Alvarez-Gaum\'e and P. Zaugg,
CERN preprint CERN-TH-6242-91.

\noindent
{\bf [14]} E. Witten,  ``Two Dimensional
Gravity and Intersection Theory on Moduli
 Space", {\it Surveys in Diff. Geom.}
{\bf 1} (1991) 243, and references therein.

R. Dijkgraaf and E. Witten, {\it Nucl. Phys.}
{\bf B342} (1990) 486.

\noindent
{\bf [15]} M. Kontsevich, ``Intersection Theory
on the Moduli Space of Curves and the
 Matrix Airy Function", Bonn preprint MPI/91-47.

\noindent
{\bf [16]} E. Witten, ``On the Kontsevich Model
and Other Models of Two Dimensional
 Gravity", IAS preprint, HEP-91/24.

\noindent
{\bf [17]} J.M. Rabin, {\it Commmun.
Math. Phys.} {\bf 137} (1991) 533.

\noindent
{\bf [18]} For details and references on
the loop equations see A.A. Migdal, {\it Phys.
Rep.}
 {\bf 102} (1983) 199.

\noindent
{\bf [19]} V. Kazakov, ``Bosonic Strings
and String Field Theories in One-Dimensional
 Target Space", lecture given at
Carg\`ese, France, May 1990, to appear in
``Random Surfaces and Quantum gravity", ed.
by O. Alvarez et.al.

\noindent
{\bf [20]} L. Alvarez-Gaum\'e,
{\it Helv. Phys. Acta} {\bf 64} (1991) 359.

\noindent
{\bf [21]} Y. Matsuo, unpublished.

\noindent
{\bf [22]} D. Bessis, C. Itzykson
 and J.-B. Zuber, {\it Adv. App. Math.} {\bf 1} (1980) 109.

\noindent
{\bf [23]} G. Moore, N. Seiberg and
M. Staudacher, {\it Nucl. Phys.} {\bf B362} (1991) 665.

\noindent
{\bf [24]} D.J. Gross and A. A. Migdal,
{\it Nucl Phys.} {\bf B340} (1990) 333.

\noindent
{\bf [25]} E. Martinec, G. Moore and
N. Seiberg, {\it Phys. Lett.} {\bf B263} (1991) 190.

\noindent
{\bf [26]} L. Alvarez-Gaum\'e and
J.L. Ma\~nes, {\it Mod. Phys. Lett.} {\bf A6} (1991)
2039.

\end